\documentclass[11pt,a4paper,notitlepage]{article}
\usepackage[mag=1000,ignoremp]{geometry}

\newtheorem{theorem}{Theorem}

\newtheorem{proposition}[theorem]{Proposition}

\input{tcilatex}
\begin{document}

\title{Competition among Large and Heterogeneous Small Firms\textbf{\thanks{%
Pan gratefully acknowledges the financial support from the Grant-in-aid for
Research Activity, Japan Society for the Promotion of Science (19K13681).}} }
\author{Lijun Pan\thanks{%
Main address: School of Economics, Nanjing University, Nanjing, China.
Affiliated address: Institute of Social and Economic Research, Osaka
University, Japan. Email: pan.lijun@hotmail.com} \\
%EndAName
\relax Nanjing University, Osaka University \and Yongjin Wang\thanks{%
School of Economics, Nankai University, Tianjin, China. Email:
yjw@nankai.edu.cn.} \\
%EndAName
\relax Nankai University}
\date{May 25, 2020}
\maketitle

\begin{abstract}
We extend the model of Parenti (2018) on large and small firms by
introducing cost heterogeneity among small firms. We propose a novel
necessary and sufficient condition for the existence of such a mixed market
structure. Furthermore, in contrast to Parenti (2018), we show that in the
presence of cost heterogeneity among small firms, trade liberalization may
raise or reduce the mass of small firms in operation.


\end{abstract}

\thispagestyle{empty} \setlength{\topmargin}{-1.0cm} \setlength{%
\textheight}{8.5in} \newpage

\pagestyle{plain}\setcounter{page}{1}

\section{Introduction}

Firm heterogeneity is prevalent in international trade. On the one hand, it
is evident that firms are heterogeneous in terms of firm size, with a few
dominant large exporters and a host of negligible small firms.\footnote{%
See Parenti (2018).} On the other hand, small firms also display
heterogeneity in the form of productivity differences.\footnote{%
See Melitz (2003) and Melitz and Ottaviano (2008).} This note attempts to
address these two types of firm heterogeneity in a coherent model and
examine the impact of trade liberalization.

We draw on the model in Parenti (2018) where large and small firms coexist,
but goes beyond that model by introducing cost heterogeneity among small
firms. In our model, large firms are treated as established incumbent
oligopolists, whereas small firms are treated as monopolistic competitors
who endogenously make entry decisions and face initial uncertainty about
their future productivity prior to entry. The heterogeneity among small
firms is in the spirit of Melitz-Ottaviano type's heterogeneity.\footnote{%
Representative works include Melitz (2003) and Melitz and Ottaviano (2008).}

First, we propose a novel necessary and sufficient condition for the
existence of such a mixed market structure in a closed economy,
accommodating the cost heterogeneity among small firms. Large and small
firms coexist if (i) the marginal cost of large firms is strictly lower than
that of the least efficient small firm in operation, and (ii) consumers'
preferences on the differentiated goods are sufficiently high, the number of
large firms is not too high, and the cost of large firms is not too low. The
former condition guarantees that large firms earn positive profits, and the
latter condition guarantees a positive mass of small firms in operation.

Second, in contrast to Parenti (2018), we show that bilateral trade
liberalization may raise the mass of small firms selling in each country.
Assuming that small firms share the same cost and do not export, Parenti
(2018) finds that trade liberalization leads to the expansion of large firms
and decreases the mass of small firms selling in each country. However, we
introduce cost heterogeneity among small firms and allow the most efficient
ones to export. Therefore, a reduction in trade cost as a result of
bilateral trade liberalization not only reduces the export cost of large
firms, but also lowers the trade cost of small firms. The former effect,
which is consistent with Parenti (2018), generates a negative impact on the
mass of small firms, whereas the latter effect generates a positive impact
on the mass of small firms. The impact of trade liberalization on the mass
of small firms selling in each country depends on these two opposing
effects. When the number and cost of large firms are sufficiently high, the
former effect dominates, and trade liberalization decreases the mass of
small firms. Otherwise, the latter effect dominates, and trade
liberalization increases the mass of small firms. We also find that trade
liberalization may raise the mass of small producers in each country.

This paper contributes to the recent studies on the mixed market structure
with large and small firms. Assuming that small firms have identical cost,
Shimomura and Thisse (2012) and Pan and Hanazono (2018) identify the
condition for the coexistence of large and small firms and examine the
impact of a large firm's entry in a closed economy. Complementary to these
works, this paper derives a new coexistence condition incorporating cost
heterogeneity among small firms and investigates the impact of trade
liberalization. Although our results can be regarded as an extension of
Parenti (2018), they open a new channel through which the most efficient
small firms could export and benefit from trade liberalization.

Our results also relate to the research on trade liberalization. In the
model with monopolistic competition and firm heterogeneity, Melitz (2003)
and Melitz and Ottaviano (2008) show that bilateral trade liberalization
raises the mass of small firms selling in each country. On the contrary, in
our model, bilateral trade liberalization may reduce the mass of small firms
due to the expansion in the export of large firms. In the Stackelberg model
where large firms are leaders and can export whereas small firms are
followers and sell only in the domestic market, Pan and Tabuchi (2019) also
find that free trade reduces the number of small firms. In contrast,
allowing for the export by efficient small firms, we find that trade
liberalization may raise the mass of small firms because the exporting small
firms also benefit from trade liberalization.

\section{Closed Economy}

Consider an economy with $L$ consumers, each supplying one unit of labor
inelastically. Each consumer consumes two types of goods, a homogeneous good
and differentiated goods. The homogeneous good is produced under constant
returns to scale at unit cost and perfect competition, which implies a unit
wage. The differentiated goods are supplied by a discrete number $N$ of
large firms $n=1,...,N$, and a continuum of small firms indexed by $i\in
\lbrack 0,M]$. Occupying a substantial market share, each large firm behaves
strategically as an oligopolist. Being negligible in the market, each small
firm behaves non-strategically as a monopolistic competitor.

\subsection{Preferences}

All consumers share the same utility function given by%
\begin{equation}
U=\alpha (\int_{0}^{M}x_{i}di+\sum_{n=1}^{N}X_{n})-\frac{\beta }{2}%
[\int_{0}^{M}(x_{i})^{2}di+\sum_{n=1}^{N}(X_{n})^{2}]-\frac{\gamma }{2}%
(\int_{0}^{M}x_{i}di+\sum_{n=1}^{N}X_{n})^{2}+x_{0},
\end{equation}%
where $x_{i}$ represents the individual consumption of the good produced by
small firm $i$, and $X_{n}$ represents the individual consumption of the
good produced by large firm $n$. The numeraire good is represented by $x_{0}$%
. The demand parameters $\alpha >0$ and $\gamma >0$ capture the degree of
substitutability between the varieties of the differentiated product and the
numeraire, and $\beta >0$ represents the degree of differentiation between
the differentiated varieties.

Assuming that consumers have positive demands for the numeraire good, the
inverse demand for small firm $i$ and large firm $n$ are respectively

\begin{equation}
p_{i}=\alpha -\beta x_{i}-\gamma \mathbf{X,}\text{ }\forall i\in \lbrack 0,M]
\label{inverse demand small}
\end{equation}%
\begin{equation}
P_{n}=\alpha -\beta X_{n}-\gamma \mathbf{X,}\text{ }n=1,...,N
\label{inverse demand big}
\end{equation}%
whenever $x_{i}>0$, $X_{n}>0$, and $\mathbf{X}$\textbf{$=$}$%
\int_{0}^{M}x_{i}di+\sum_{n=1}^{N}X_{n}$ is the individual total consumption
of the differentiated good$.$

Inverse demands (\ref{inverse demand small}) and (\ref{inverse demand big})
can be inverted to the demands for small firm $i$ and large firm $n$:%
\begin{equation}
q_{i}=Lx_{i}=\frac{\alpha L}{\beta +\gamma (M+N)}-\frac{L}{\beta }p_{i}+%
\frac{L}{\beta }\frac{\gamma }{\beta +\gamma (M+N)}\mathbf{P,}\text{ }%
\forall i\in \lbrack 0,M]  \label{demand small}
\end{equation}%
\begin{equation}
Q_{n}=LX_{n}=\frac{\alpha L}{\beta +\gamma (M+N)}-\frac{L}{\beta }P_{n}+%
\frac{L}{\beta }\frac{\gamma }{\beta +\gamma (M+N)}\mathbf{P,}\text{ }%
n=1,...,N  \label{demand big}
\end{equation}%
where $\mathbf{P}=\int_{0}^{M}p_{i}di+\sum_{n=1}^{N}P_{n}$ is the aggregate
price. To ensure a positive consumption, i.e., $q_{i}\geq 0$ and $Q_{n}\geq
0 $, the prices of large and small firms should satisfy 
\begin{equation}
p_{i},P_{n}\leq \frac{1}{\beta +\gamma (M+N)}(\alpha \beta +\gamma \mathbf{P}%
)=p_{\max },  \label{pmax}
\end{equation}%
where $p_{\max }$ represents the price at which demand for the good is
driven to zero.

\subsection{Firms}

We consider a 2-stage game. In the first stage, small firms enter the
market. In the second stage, large and small firms who have entered the
market compete in price.

\paragraph{Small Firms}

Entry in the differentiated product sector is costly as each firm incurs
product development and production start-up costs. Subsequent production
exhibits constant returns to scale at a marginal cost $c$. Following Melitz
and Ottaviano (2008), Research and development yield uncertain outcomes for $%
c,$ and firms learn about this cost level only after making the irreversible
investment $f_{E}$ required for entry. We assume that the marginal cost of
each small firm is a draw from a common and known distribution $G(c)$ with
support $[0,c_{M}].$ Since the entry cost is sunk, firms that can cover
their marginal cost survive and produce. All other firms exit the industry.

Substituting the demand function (\ref{demand small}), the gross profit of
small firm $i$ with marginal cost $c_{i}$ can be expressed as%
\begin{equation}
\pi _{i}=(p_{i}-c_{i})(\frac{\alpha L}{\beta +\gamma (M+N)}-\frac{L}{\beta }%
p_{i}+\frac{L}{\beta }\frac{\gamma }{\beta +\gamma (M+N)}\mathbf{P}),\text{ }%
\forall i\in \lbrack 0,M].  \label{pis}
\end{equation}

\paragraph{Large Firms}

Large firms are established incumbents and produce at marginal cost $C$.
Substituting the demand function (\ref{demand big}), the profit of large
firm $n$ can be expressed as%
\begin{equation}
\Pi _{n}=(P_{n}-C)(\frac{\alpha L}{\beta +\gamma (M+N)}-\frac{L}{\beta }%
P_{n}+\frac{L}{\beta }\frac{\gamma }{\beta +\gamma (M+N)}\mathbf{P}),\text{ }%
n=1,...,N.  \label{piL}
\end{equation}

\subsection{The Equilibrium Analysis}

\subsubsection{\textbf{Stage 2}}

\paragraph{Small Firms}

Small firm $i$ maximizes its profit with respect to its price $p_{i}$,
treating the aggregate price $\mathbf{P}$ as given since its market share is
negligible. The profit maximizing price $p_{i}$ and quantity $q_{i}$ then
satisfy%
\begin{equation}
q_{i}(c_{i})=\frac{L}{\beta }[p_{i}(c_{i})-c_{i}],  \label{qs(ps)}
\end{equation}

If the profit maximizing price $p_{i}$ is above the price bound $p_{\max }$
from (\ref{pmax}), then small firm $i$ exits. Let $c_{D}$ denote the cost of
the small firm who is indifferent about remaining in the industry. This firm
earns zero profit as its cost is equal to its marginal cost, that is, $%
p(c_{D})=c_{D}=p_{\max }$, and its demand level $q(c_{D})$ is driven to $0$.
We assume that $c_{M}$ is high enough to be above $c_{D}$, so that some
firms with cost draws between these two levels exit. All firms with cost
below $c_{D}$ earn positive profits (gross of the entry cost) and remain in
the industry. Substituting (\ref{pmax}), and knowing that $p_{\max }=c_{D}$,
the first order condition of a small firm with $c<c_{D}$ yields its optimal
price $p(c)$ and quantity $q(c)$:%
\begin{equation}
p(c)=\frac{1}{2}(c_{D}+c),  \label{ps(c)}
\end{equation}%
\begin{equation}
q(c)=\frac{L}{2\beta }(c_{D}-c),  \label{qs(c)}
\end{equation}%
and the profit can be expressed by%
\begin{equation}
\pi (c)=\frac{L}{4\beta }(c_{D}-c)^{2}.  \label{pis(c)}
\end{equation}

\paragraph{Large Firms}

Different from a small firm, who treats the aggregate price $\mathbf{P}$
parametrically, a large firm behaves strategically and internalizes its
impact on $\mathbf{P}$. Large firm $n$ maximizes its profit given by (\ref%
{piL}), yielding the optimal price 
\begin{equation}
P_{n}=P=\frac{c_{D}+(1-\Theta )C}{2-\Theta },\text{ }n=1,...,N,  \label{PL}
\end{equation}%
where $\Theta =\gamma /[\beta +\gamma (M+N)]\in (0,1)$ represents the
internalization by the large firm$.$ Due to the internalization, a large
firm's price depends on the threshold cost $c_{D}$ of small firms and the
mass $M$ of small firms, which are endogenously determined by the free entry
condition in the first stage$.$ Moreover, incurring the same marginal cost $%
C $, large firms set the same price $P$.

Similar to a small firm, a large firm stops operation if and only if the
profit maximizing price $P$ is above the price bound $p_{\max }$. Since $%
p_{\max }=c_{D}$, the necessary and sufficient condition for the existence
of large firms is $C<c_{D}$.\footnote{%
We assume that large firms incur zero fixed cost. If the fixed cost is
positive, then the condition for the existence of large firms should be
stricter.}

\subsubsection{\textbf{Stage 1}}

Prior to entry, the expected profit of a small firm\ is $\int_{0}^{c_{D}}\pi
(c)dG(c)-f_{E}.$ Substituting (\ref{pis(c)}), the free entry condition can
be rewritten as%
\begin{equation}
\frac{L}{4\beta }\int_{0}^{c_{D}}(c_{D}-c)^{2}dG(c)=f_{E},
\label{free entry}
\end{equation}%
which determines the cost cut-off $c_{D}$.

To obtain tractable results, we assume the Pareto parametrization for the
cost draws of small firms, i.e., 
\begin{equation}
G(c)\,=(\frac{c}{c_{M}})^{k},\text{ }c\in \lbrack 0,c_{M}].
\label{pareto cost}
\end{equation}

Given this parametrization, the cut-off cost level $c_{D}$ determined by (%
\ref{free entry}) is then 
\begin{equation}
c_{D}^{\ast }=(\frac{\beta \phi }{L})^{\frac{1}{k+2}},  \label{cD}
\end{equation}%
where $\phi =2(k+1)(k+2)(c_{M})^{k}f_{E}$ is the technology index.

Substituting the optimal prices of small and large firms given by (\ref%
{ps(c)}) and (\ref{PL}), and using that $p_{\max }=c_{D}$, (\ref{pmax}) can
be expressed as%
\begin{equation}
\frac{(\alpha -c_{D}^{\ast })\beta }{\gamma }=\frac{Mc_{D}^{\ast }}{2(k+1)}%
+N(c_{D}^{\ast }-C)\frac{1-\Theta (M)}{2-\Theta (M)},  \label{M*}
\end{equation}%
In equation (\ref{M*}), the LHS is constant and positive, whereas the RHS
increases with $M$. Let $M^{\ast }$ be the equilibrium mass of small firms.
To ensure that $M^{\ast }>0$, the RHS should be smaller than the LHS when $%
M=0$. Therefore, equation (\ref{M*}) uniquely determines a positive mass $%
M^{\ast }$ of small firms in operation if and only if 
\[
\frac{(\alpha -c_{D}^{\ast })\beta }{\gamma }>N(c_{D}^{\ast }-C)\frac{\beta
+\gamma (N-1)}{2\beta +\gamma (2N-1)}.
\]

As shown earlier, large firms earn a positive profit if and only if $C<c_{D}$%
. Substituting the equilibrium cost cut-off of small firms in (\ref{cD}), we
establish the condition for the coexistence of large and small firms in
Proposition 1.

\begin{proposition}
There exists a unique mixed market equilibrium where large and small firms
coexist if and only if (i) $C<(\beta \phi /L)^{\frac{1}{k+2}}$ and (ii) $%
[\alpha -(\beta \phi /L)^{\frac{1}{k+2}}]\beta >\gamma N[(\beta \phi /L)^{%
\frac{1}{k+2}}-C][\beta +\gamma (N-1)]/[2\beta +\gamma (2N-1)]$.
\end{proposition}

The first condition, which guarantees that large firms earn positive
profits, requires that the cost of large firms be smaller than the cost
cutoff of small firms, which is satisfied if the technology index $\phi $ is
sufficiently high and the market size $L$ is not too large. The second
condition, which guarantees a positive mass of small firms, is satisfied if
the consumer's preference for the differentiated good $\alpha $ is
sufficiently high, the cost $C$ of large firms is not too low in comparison
with the cost cutoff $(\beta \phi /L)^{\frac{1}{k+2}}$ of small firms, and
the number $N$ of large firms is not too large.

In models with the coexistence of large and small firms, see, e.g.,
Shimomura and Thisse (2012), Parenti (2018), and Pan and Hanazono (2018), it
is commonly assumed that small firms share the same technology. When large
firms are single-product and facing zero fixed cost, their assumption
implies large firms operate if and only if $C<c+2\sqrt{\beta f_{E}}$.
However, when we allow the small firms to be heterogeneous and facing cost
uncertainty prior to entry, Proposition 1 indicates that large firms should
be more efficient than the least productive small firm to survive.

\section{Open Economy}

In the previous section, we considered a closed-economy model. In this
section, we extend the model to an open economy, and examine the impact of
trade liberalization. Following Parenti (2018), we consider two symmetric
countries, $H$ and $F$, each with the same market size and distribution of
firms. Specifically, each country has $L$ consumers who share the same
preferences, leading to the inverse demand functions (\ref{inverse demand
small}) and (\ref{inverse demand big}). In addition, in each country, there
are $N$ large firms with marginal cost $C$, and small firms share the same
cost distribution given by (\ref{pareto cost}). The two markets are
segmented, although firms can produce in one market and sell in the other at
a trade cost $\tau $. We assume that large and small firms exist and export
in both countries, and we will identify the condition in the analysis.

The aggregate price in country $h$ is then defined by%
\[
\mathbf{P}^{h}=\int_{0}^{M^{h}}p_{i}di+NP_{D}^{h}+NP_{X}^{f},\text{ }h,f=H,F,%
\text{ }h\neq f, 
\]%
where $M^{h}$ is the mass of small firms selling in country $h,$ including
both domestic small firms and foreign small exporters, $P_{D}^{h}$
represents the price of domestic large firms, and $P_{X}^{f}$ represents the
delivered price of foreign large firms.

\bigskip The price threshold for positive demand in country $h$ is then
given by 
\begin{equation}
p_{\max }^{h}=\frac{1}{\beta +\gamma (M^{h}+2N)}(\alpha \beta +\gamma 
\mathbf{P}^{h}).\text{\ }  \label{pmax l}
\end{equation}

\subsection{The Equilibrium Analysis}

\subsubsection{Stage 2}

\paragraph{Small Firms}

Let $p_{D}^{h}(c)$ and $q_{D}^{h}(c)$ represent the domestic levels of
profit maximizing price and quantity sold for a small firm producing in
country $h$ with cost $c$. Such a small firm may also export output $%
q_{X}^{h}(c)$ at a delivered price $p_{X}^{h}(c).$ The delivered cost of a
unit with cost $c$ to the foreign country is $\tau c,$ where $\tau >1$ and
is the same across all exporting firms. Since the markets are segmented,
small firms independently maximize the profits in the domestic and foreign
markets. Let $\pi _{D}^{h}(c)=[p_{D}^{h}(c)-c]q_{D}^{h}(c)$ and $\pi
_{X}^{h}(c)=[p_{X}^{h}(c)-\tau c]q_{X}^{h}(c)$ denote the maximized value of
these profits as a function of the firm's marginal cost $c$.

Analogously to (\ref{qs(ps)}), the profit maximizing prices and quantities
must satisfy: $q_{D}^{h}(c)=(L/\beta )[p_{D}^{h}(c)-c]$ and $%
q_{X}^{h}(c)=(L/\beta )[p_{X}^{h}(c)-\tau c].$ As was the case in the closed
economy, only the small firms that earn non-negative profits in a market
(domestic or export) will choose to sell in that market. This leads to
similar cost cut-off rules for small firms selling in either market. Let $%
c_{D}^{h}$ denote the upper bound cost for small firms selling in their
domestic market, and let $c_{X}^{h}$ denote the upper bound cost for small
exporters from $h$ to $f$. These cutoffs must then satisfy:%
\begin{eqnarray}
c_{D}^{h} &=&\sup \{c:\pi _{D}^{h}(c)>0\}=p_{\max }^{h},  \label{cutoff pmax}
\\
c_{X}^{h} &=&\sup \{c:\pi _{X}^{h}(c)>0\}=\frac{p_{\max }^{f}}{\tau }. 
\nonumber
\end{eqnarray}

This implies $c_{X}^{f}=c_{D}^{h}/\tau $: trade barriers make it harder for
small firms to break even in the foreign market than in the domestic market.

As was the case in the closed economy, the optimal prices and quantities are
also functions of the cost cutoffs:%
\begin{eqnarray}
p_{D}^{h}(c) &=&\frac{1}{2}(c_{D}^{h}+c),\text{ }q_{D}^{h}(c)=\frac{L}{%
2\beta }(c_{D}^{h}-c),  \label{p(c) open} \\
p_{X}^{h}(c) &=&\frac{\tau }{2}(c_{X}^{h}+c),\text{ }q_{X}^{h}(c)=\frac{L}{%
2\beta }\tau (c_{X}^{h}-c),  \nonumber
\end{eqnarray}%
which yield the maximized profit levels:%
\begin{eqnarray}
\pi _{D}^{h}(c) &=&\frac{L}{4\beta }(c_{D}^{h}-c)^{2},  \label{pi(c) open} \\
\pi _{X}^{h}(c) &=&\frac{L}{4\beta }\tau ^{2}(c_{X}^{h}-c)^{2}.  \nonumber
\end{eqnarray}

\paragraph{Large Firms}

Similar to the closed economy, we assume that all the large firms share the
same marginal cost $C$ in production, and the delivered cost of a unit to
the foreign country is $\tau C$. Let $P_{D}^{h}$ and $Q_{D}^{h}$ represent
the domestic levels of profit maximizing price and quantity sold for a large
firm producing in country $h$. The large firm also exports output $Q_{X}^{h}$
at a delivered price $P_{X}^{h}.$ Analogously to (\ref{PL}), the profit
maximizing domestic and export prices are%
\begin{eqnarray}
P_{D}^{h} &=&\frac{c_{D}^{h}+(1-\Theta ^{h})C}{2-\Theta ^{h}},
\label{P open} \\
P_{X}^{h} &=&\tau \frac{c_{X}^{h}+(1-\Theta ^{f})C}{2-\Theta ^{f}}, 
\nonumber
\end{eqnarray}%
where $\Theta ^{h}=\gamma /[\beta +\gamma (M^{h}+2N)]$ and $\Theta
^{f}=\gamma /[\beta +\gamma (M^{f}+2N)]$. A large firm in country $h$ stops
selling in the domestic market if and only if $P_{D}^{h}$ is above the price
bound $p_{\max }^{h}$. Since $p_{\max }^{h}=c_{D}^{h}$, the necessary and
sufficient condition for the domestic operation of large firms is $%
C<c_{D}^{h}$. Furthermore, it stops exporting if and only if $P_{X}^{h}$ is
above the price bound $p_{\max }^{f}$ (where $f\neq h$). Since $p_{\max
}^{f}=\tau c_{X}^{h}$, the necessary and sufficient condition for the
existence of exporting large firms is $C<c_{X}^{h}.$ These two conditions
indicate that large firms share the same cost cutoffs with small firms in
both domestic and foreign markets.

Therefore, large firms in both countries export if and only if $C<\min
\{c_{X}^{H},c_{X}^{F}\}.$

\subsubsection{Stage 1}

Entry of small firms is unrestricted in both countries. Small firms choose a
production location prior to entry and pay the sunk entry cost. We assume
that both countries share the same technology for small firms -- referenced
by the entry cost $f_{E}$ and cost distribution $G(c)$. Therefore, the free
entry condition\ is expressed as%
\[
\frac{L}{4\beta }\int_{0}^{c_{D}^{h}}\pi
_{D}^{h}(c)dG(c)+\int_{0}^{c_{X}^{h}}\pi _{X}^{h}(c)dG(c)=f_{E}. 
\]

\bigskip With Pareto parametrization for the cost draws $G(c)$ in both
countries, the free entry condition can be rewritten as 
\[
(c_{D}^{h})^{k+2}+\tau ^{2}(c_{X}^{h})^{k+2}=\frac{\beta \phi }{L}. 
\]

Since $c_{X}^{f}=c_{D}^{h}/\tau $, the free entry condition (\ref{free entry
open}) can be expressed as 
\begin{equation}
(c_{D}^{h})^{k+2}+\rho (c_{D}^{f})^{k+2}=\frac{\beta \phi }{L},
\label{free entry open}
\end{equation}%
where $\rho =(\tau )^{-k}\in (0,1)$ is an inverse measure of trade costs
(the \textquotedblleft freeness\textquotedblright\ of trade). This system
can be solved for the cutoffs in both countries:%
\begin{equation}
c_{D}^{h\ast }=c_{D}^{f\ast }=c_{D}^{\ast }=[\frac{\beta \phi }{L(1+\rho )}%
]^{1/(k+2)},  \label{cD open}
\end{equation}%
by which the average cost of small firms selling in each country is $\bar{c}%
_{D}^{\ast }=kc_{D}^{\ast }/(k+1).$

By (\ref{pmax l}) and (\ref{P open}), the aggregate price in country $h$ is
given by

\[
\mathbf{P}^{h}=\frac{M^{h}(2k+1)}{2(k+1)}c_{D}^{\ast }+N\frac{c_{D}^{\ast
}+(1-\Theta ^{h})C}{2-\Theta ^{h}}+N\frac{c_{D}^{\ast }+(1-\Theta ^{f})\tau C%
}{2-\Theta ^{f}},
\]%
Substituting $\mathbf{P}^{h}$, (\ref{pmax l}) can be expressed as%
\begin{equation}
\frac{(\alpha -c_{D}^{\ast })\beta }{\gamma }=\frac{M^{h}}{2}\frac{%
c_{D}^{\ast }}{k+1}+N(c_{D}^{\ast }-C)\frac{1-\Theta ^{h}}{2-\Theta ^{h}}%
+N(c_{D}^{\ast }-\tau C)\frac{1-\Theta ^{h}}{2-\Theta ^{h}}.
\label{Ml(pmax) open pareto}
\end{equation}%
which uniquely determines the mass of small firms in country $h$ because the
RHS strictly increases with $M^{h}$. (\ref{Ml(pmax) open pareto}) also
implies that $M^{f}=M^{h}=M$ and hence $\Theta ^{f}=\Theta ^{h}=\Theta .$
Therefore, (\ref{Ml(pmax) open pareto}) could be reduced to%
\begin{equation}
\frac{(\alpha -c_{D}^{\ast })\beta }{\gamma }=\frac{M}{2}\frac{c_{D}^{\ast }%
}{k+1}+\frac{1-\Theta (M)}{2-\Theta (M)}N[(c_{D}^{\ast }-C)+(c_{D}^{\ast
}-\tau C)],  \label{Ml(pmax) open pareto symmetric}
\end{equation}%
which uniquely determines the equilibrium mass $M^{\ast }$ of small firms
selling in each country. To ensure that $M^{\ast }>0$, we assume that $%
(\alpha -c_{D}^{\ast })\beta /\gamma >N[(c_{D}^{\ast }-C)+(c_{D}^{\ast
}-\tau C)][\beta +\gamma (N-1)]/[2\beta +\gamma (2N-1)].$

\subsection{\protect\bigskip Trade Liberalization}

Based on the above equilibrium analysis, now we examine how bilateral trade
liberalization impacts the mass of small firms.

First, (\ref{cD open}) indicates that a reduction in trade cost as a result
of bilateral trade liberalization decreases the cost cut-off of small firms,
that is, $dc_{D}^{\ast }/d\tau >0$.

Now we examine the impact of trade liberalization on the mass of small firms
selling in each country, including both domestic small producers and foreign
small exporters. By (\ref{Ml(pmax) open pareto symmetric}), we have%
\[
\frac{dM}{d\tau }=-\frac{[\frac{\alpha \beta }{\gamma }+(1+\tau )NC]\frac{k}{%
k+2}\frac{\rho }{\rho +1}\frac{1}{\tau }-\frac{1-\Theta }{2-\Theta }NC}{%
\frac{c_{D}^{\ast }}{2(k+1)}+(\frac{\Theta }{2-\Theta })^{2}[(c_{D}^{\ast
}-C)+(c_{D}^{\ast }-\tau C)]N}, 
\]%
which is negative if and only if%
\begin{equation}
\frac{2-\Theta (M^{\ast })}{1-\Theta (M^{\ast })}\frac{\alpha \beta }{\gamma 
}>(\frac{2\rho +k+2}{k\rho }\tau -1)NC,  \label{M condition}
\end{equation}%
where $M^{\ast }$ is the equilibrium mass of small firms determined by (\ref%
{Ml(pmax) open pareto symmetric}).

Proposition 2 establishes the impact of bilateral trade liberalization on
the mass of small firms selling in each country.

\begin{proposition}
A decrease in trade costs as a result of bilateral trade liberalization
increases the mass of small firms selling in each country if $(\alpha \beta
/\gamma )[2-\Theta (M^{\ast })]/[1-\Theta (M^{\ast })]>[(2\rho +k+2)\tau
/(k\rho )-1]NC$, and decreases the mass of small firms selling in each
country otherwise.
\end{proposition}

A reduction in trade cost generates \textit{two opposing effects} on the
mass of small firms selling in each country. First, trade liberalization
reduces the export cost and hence enables more small firms to export, which
potentially increases the mass of small firms. Second, a decrease in trade
cost also generates cost savings on the exported goods of large firms, which
induces large firms to expand export. This generates competitive pressure on
small firms and may reduce the mass of surviving small firms. As marginal
cost $C$ rises, the cost savings become stronger and lead to more aggressive
exporting behavior by large firms. Furthermore, the aggregation of such
aggressive exports increases with the number $N$ of large firms. Therefore,
condition (\ref{M condition}) holds as long as the number $N$ or marginal
cost $C$ of large firms is not too high.

Last, we show that bilateral trade liberalization may also raise the mass of
small producers in each country. By symmetry, in each country, there are $%
M_{E}G(c_{D})$ domestic small producers and $M_{E}G(c_{X})$ foreign small
exporters. Thus, $M_{E}G(c_{D})+M_{E}G(c_{D}/\tau )=M$, by which the mass of
small entrants in each country can be expressed as $%
M_{E}=M(c_{M}/c_{D})^{k}/(1+\rho )$. The mass of small producers in each
country is then expressed by 
\[
M_{D}=M_{E}G(c_{D})=\frac{M}{1+\rho }. 
\]%
The impact of a reduction in trade cost on the mass of small producers is
then given by%
\[
\frac{dM_{D}}{d\tau }=\frac{1}{1+\rho }[\frac{dM}{d\tau }+\frac{k\rho M}{%
(1+\rho )\tau }]. 
\]

A reduction in trade cost impacts the mass $M_{D}$ of domestic small
producers through (a) the change in the mass $M$ of small firms selling in
each country and (b) the reallocation between domestic producers and foreign
exporters within the group of small firms. The change in the mass of small
firms selling in each country, according to Proposition 2, is positive as
long as thenumber and marginal cost of large firms are not too high.
Furthermore, trade liberalization invites the most productive small firms to
increase export, which makes competition tougher and may force some of the
least productive small producers to exit. This generates a negative impact
on the mass of domestic small producers. Therefore, if the rise in the total
mass of small firms selling in each country outweighs the negative impact
from the export expansion by the most productive small firms, then trade
liberalization would also raise the mass of small producers in each country.
Otherwise, trade liberalization reduces the mass of small producers.
Proposition 3 establishes this result with the precise condition.

\begin{proposition}
A decrease in trade costs as a result of bilateral trade liberalization
increases the mass of small producers if 
\[
\frac{\alpha \beta /\gamma +(1+\tau )NC}{k+2}>\frac{M^{\ast }c_{D}^{\ast }}{%
2(k+1)}+M^{\ast }N[2c_{D}^{\ast }-(1+\tau )C][\frac{\Theta (M^{\ast })}{%
2-\Theta (M^{\ast })}]^{2}+NC\frac{(1+\rho )\tau }{\rho k}\frac{1-\Theta
(M^{\ast })}{2-\Theta (M^{\ast })}, 
\]%
and decreases the mass of small producers otherwise.
\end{proposition}

Here $c_{D}^{\ast }$ and $M^{\ast }$ are the equilibrium values determined
in (\ref{cD open}) and (\ref{Ml(pmax) open pareto symmetric}). Owing to the
expansion of small exporters as a result of trade liberalization, the range
of parameters where trade liberalization increases the mass of small
producers narrows down.\footnote{%
We identify both the case when $dM_{D}^{\ast }/d\tau >0$ and that when $%
dM_{D}^{\ast }/d\tau <0$ by numerical analysis. For instance, when $\alpha
=0.6$, $\beta =\gamma =1,$ $N=1$, $C=0.01$, $L=100$, $f_{E}=1$, and $\tau
=1.5$, $dM_{D}^{\ast }/d\tau <0$; when $\alpha =2.4$, $\beta =\gamma =1,$ $%
N=2$, $C=0.02$, $L=100$, $f_{E}=1$, and $\tau =1.5$, $dM_{D}^{\ast }/d\tau
>0 $.}

We conclude by highlighting the distinction between our results and those in
Parenti (2018). Take Proposition 2 for instance. Assuming that small firms
share the same cost and do not export, Parenti (2018) finds that trade
liberalization decreases the mass of small firms. In Proposition 2, we go a
step further by introducing cost heterogeneity among small firms and
allowing a portion of small firms to export, identifying two opposing
effects of trade liberalization on the mass of small firms. The second
effect, which reduces the mass of small firms, is consistent with Parenti
(2018), whereas the first effect is novel here. The insights of Proposition
2 extend to Proposition 3.

\end{document}